\documentclass[aps,pra,twocolumn,showpacs,showkeys,superscriptaddress,reprint,floatfix]{revtex4-1} 
\usepackage{graphics} 
\usepackage{graphicx} 
\usepackage{amsmath} 
\usepackage{amssymb}
\usepackage{color}
\usepackage[caption=false]{subfig}

\begin{document} 

\title{Phase diagram of the SU$(3)$ Fermi-Hubbard model with next-neighbor interactions} 
\author{A. P\'erez-Romero}
\affiliation{Departamento de F\'{\i}sica, Universidad Nacional de Colombia, A. A. 5997 Bogot\'a, Colombia.}
\author{R. Franco} 
\affiliation{Departamento de F\'{\i}sica, Universidad Nacional de Colombia, A. A. 5997 Bogot\'a, Colombia.}
\author{J. Silva-Valencia} 
\email{jsilvav@unal.edu.co} 
\affiliation{Departamento de F\'{\i}sica, Universidad Nacional de Colombia, A. A. 5997 Bogot\'a, Colombia.} 

\date{\today} 

\begin{abstract}
We explore the zero-temperature phase diagram of a one-dimensional gas composed of three-color fermions, which interact locally and  with their next neighbors. Using the density matrix renormalization group method and considering one-third filling, we characterize the ground state for several values of the parameters, finding diverse phases, namely: phase separation, spin density wave, pairing phase, a metallic phase, two different charge-density waves, and a non-separable state with modulation of charge. We show that the von Neumann block entropy and the fidelity susceptibility are useful for estimating the borders between the phases.
\end{abstract} 


\maketitle 

\section{\label{sec1}Introduction}
Ultracold atom setups have become an experimental platform for studying many-body physics and other phenomena. In fully-control environments, researchers have created exotic lattices and extended some concepts and interaction processes~\cite{Lewenstein-AP07,IBloch-RMP08,IBloch-NP12}. 
Confining atoms that possess a large spin degeneracy such as $^{6}Li$, $^{87}Sr$, and $^{173}Yb$, among others, it has been possible to achieve a degenerate gas of carriers with several internal degrees of freedom ($N>2$), thus realizing the exotic $SU(N)$ systems, which have exhibited several new physical properties and possibilities compared to their well-known $SU(2)$ counterpart~\cite{Cazalilla-RPP14,Capponi-AP16,Dobrzyniecki-ArX2020}.\par 
Some remarkable experimental results related to a degenerate gas of atoms with several hyperfine states are:  the realization of  a $SU(6)$ Mott insulator phase with $^{173} Yb$ atoms~\cite{Taie-NP12,Hofrichter-PRX16},  evidence that the antiferromagnetic correlation is enhanced for the $SU(4)$-spin system compared with $SU(2)$ as a consequence of a Pomeranchuk cooling effect~\cite{HOzawa-PRL18}, measurement of spin-exchanging contact interactions in a two-orbital $SU(N)$ gas~\cite{Pagano-PRL15,Zhang-S14,Riegger-PRL18,KOno-PRA19}, and evidence of bosonization of $SU(N)$ fermions in one and three dimensions~\cite{Pagano-NP14,BSong-ArX2020}, among others.\par 
\begin{figure}[t] 
\includegraphics[width=20.5pc]{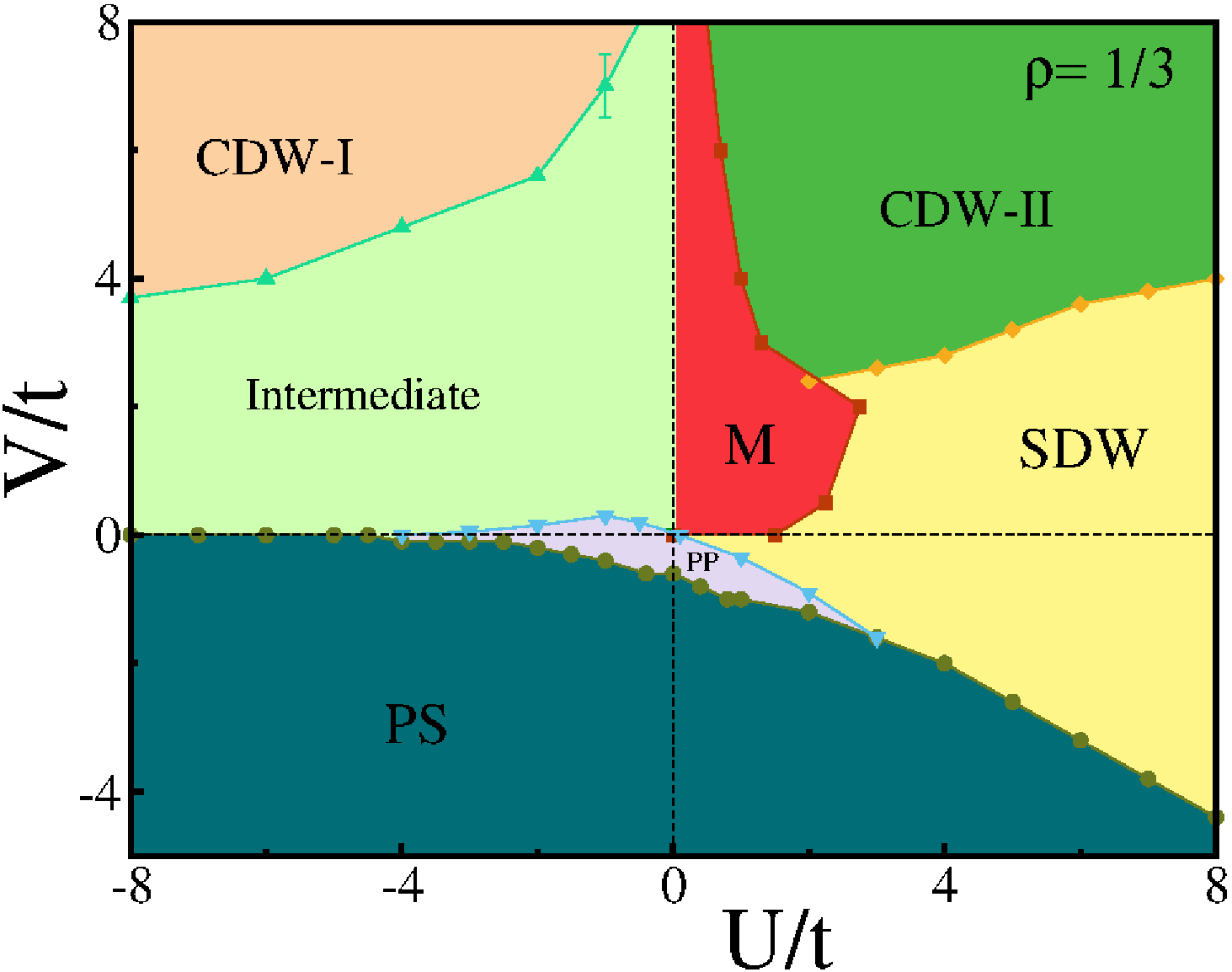} 
\caption{\label{fig1} Phase diagram of the one-third extended SU$(3)$ Hubbard model in one dimension. The parameter space is divided into the following phases: phase separation (PS), pairing phase (PP), spin-density wave (SDW), metallic phase (M), an intermeadiate phase, and two  charge-density wave phases (see text for descriptions). The points are borders obtained with DMRG, and the lines are visual guides.}
\end{figure} 
The decoupling of nuclear spin from electronic angular momentum in some atoms allows describing $SU(N)$ degenerate gases in terms of a $SU(N>2)$ Fermi-Hubbard model, generalizing the well-known model for solids~\cite{Gorshkov-NP10,Cazalilla-RPP14}. Again, the physical properties are determined by a kinetic term and a local interaction, but the key here is the number of internal degrees of freedom, which generates new characteristics~\cite{Capponi-AP16}. We point out that for $N>2$, a gapless-to-Mott insulator transition for a finite local repulsion takes place for $1/N$ filling, contrary to the exact result for $N=2$~\cite{Assaraf-PRB99,Buchta-PRB07,Manmana-PRA11}. Recently, the phase diagram of the repulsive and attractive $SU(3)$ Hubbard model on a honeycomb lattice was calculated, finding diverse ground states and several quantum phase transitions~\cite{Chung-PRB19,HXu-ArX2019}.\par 
Graphene~\cite{Wehling-PRL11}, aromatic molecules~\cite{Pariser-JCP53,Pople-PPS55}, and Fabre charge-transfer salts~\cite{Jacko-PRB13} are examples of materials that require a next-neighbor interaction term in their description, leading to the extended $SU(2)$ Hubbard model, which has been studied extensively~\cite{Emery-CBook79,Hirsch-PRL84,HQLin-PRB86,Cannon-PRB90,Cannon-PRB91,XZYan-PHB93,Sano-JPSJ94,VanDongen-PRB94,MNakamura-PRB00,Tsuchiizu-PRL02,Jeckelmann-PRL02,Sengupta-PRB02,GPZhang-PRB03,Sandvik-PRL04,YZZhang-PRL04,KMTam-PRL06,Ejima-PRL07,Glocke-PRB07,Mancini-PRE08}. Their phase diagram is composed of the phases singlet and triplet superconductor, phase separation (PS), spin density wave (SDW), charge density wave (CDW), and bond order wave (BOW).\par  
Although next-neighbor interactions are little relevant in some ultracold atom setups, for dipolar gases these interactions  compete with short-range ones, leading to novel types of order, such as the formation of dipolar quantum crystals~\cite{Sieberer-PRA11} and topological superfluidity~\cite{Levinsen-PRA11}.  Recently, it was shown that relevant non-local interactions can be achieved in Rydberg atom  setups, allowing us to study extended Hubbard models for fermions and bosons~\cite{GuardadoS-Arxiv20,Barbier-Arxiv21}. The above scenario motivated us to consider the effect of next-neighbor interactions in a three-color fermion system. Our main result is the phase diagram of a one-third-filled model, shown in Fig.~\ref{fig1}. Here, SDW, pairing, and phase separation phases appear; however, the most relevant finding is the emergence of a metallic phase, two different charge-density wave phases and an intermeadiate phase.\par 
The rest of this paper is organized as follows: In Sec.~\ref{sec2}, we explain the extended $SU(3)$ Hubbard Hamiltonian and some quantities used to characterize the phases. An exploration of the ground state is discussed in Sec.~\ref{sec3}, as well as the criteria used to establish the boundaries between the phases. Finally, we summarize our principal results in Sec. ~\ref{sec4}.
\section{\label{sec2} Model}
A one-dimensional system composed of 3-flavor fermions that can interact locally and with their next-neighbors is described by the following Hamiltonian:
\begin{eqnarray}
\label{Hamil}
H&=&-t \sum^{L-1}_{\langle i,j \rangle} \sum_\tau \left(\hat{c}^{\dagger} _{i,\tau}\hat{c}_{j,\tau}+ H.c.\right)+V \sum^{L-1}_{\langle i,j \rangle} \hat{n}_i \hat{n}_j \nonumber \\
&  &\quad \quad+\frac{U}{2} \sum^{L}_i \sum_{\tau,\tau ' \neq \tau}\hat{n}_{i,\tau}\hat{n}_{i,\tau '},
\end{eqnarray}
\noindent where $L$ is the number of sites in the chain, $\hat{c}^{\dagger} _{i,\tau}$ ($\hat{c}_{j,\tau}$) is the fermionic creation (annihilation) operator that creates (destroys) a particle at site $i$ with a color $\tau$, the sum symbol, $\sum_\tau$,  runs over the three possible colors ($\alpha$, $\beta$, and $\gamma$), $t$ is the hopping integral between neighbor sites, and $U$ and $V$ parameterize the on-site and  nearest-neighbor interactions, respectively. The subscript $\langle i,j \rangle$ means that in the sum, only the next-neighbor sites were considered. And finally, $\hat{n}_{i} \equiv \sum_\tau \hat{n}_{i,\tau}$ ($\hat{n}_{i,\tau} \equiv \hat{c}^{\dagger} _{i,\tau}\hat{c}_{i,\tau}$) is the particle number operator at site $i$ (with color $\tau$). We fix our energy scale taking $t=1$ in the Hamiltonian.\par 
We recover the one-dimensional $SU(3)$ Hubbard model with $V=0$, for which a quantum phase transition for a finite value of $U$ takes place for a global density $\rho=1/3$. Namely, it was found that $U_c\approx 2.2t$ by Assaraf \textit{et al.}~\cite{Assaraf-PRB99} and the most recent calculation yielded $U_c\approx 1.5t$~\cite{Manmana-PRA11}, which was corroborated by us.\par 
Charge and spin gaps are useful quantities for characterizing the ground state of a system, which for a finite size are given by:  
\begin{eqnarray}\label{Cgap}
\Delta_{C}(L)&=& E(N_\alpha+1,N_\beta,N_\gamma,L)+ E(N_\alpha-1,N_\beta,N_\gamma,L) \nonumber \\
& &-2E(N_\alpha,N_\beta,N_\gamma,L),
\end{eqnarray}
\noindent and
\begin{equation}\label{Sgap}
\Delta_{S}(L)= E(N_\alpha-1,N_\beta+1,N_\gamma,L)-E(N_\alpha,N_\beta,N_\gamma,L),
\end{equation}
\noindent where $E(N_\alpha,N_\beta,N_\gamma,L)$ is the ground-state energy for a chain of size $L$, with $N_\alpha$, $N_\beta$, and $N_\gamma$ fermions. This ground-state energy was calculated using the  density matrix renormalization group (DMRG) method ~\cite{White-PRL92,Hallberg-AP06,Schollwock-RMP05,Schollwock-AP11}. The numerical calculations reported here were obtained with our own code and then corroborated and extended with the Itensor library~\cite{itensor}. For the latter with open boundary conditions, we performed 40 sweeps, keeping a discarded weight of $\sim10^{-9}$, obtaining an  absolute convergence of the energy of $10^{-5}$ or less. The maximum number of states kept per block was $3000$.\par 
Spin and charge structure factors obtained from density-density correlations can be used to differentiate different phases and can be written like this:
\begin{equation}\label{SSF}
S(k)=\frac{1}{L}\sum_{l,m=1}^{L}e^{ik(l-m)}\Bigl(\langle \hat{n}_{l,\tau} \hat{n}_{m,\tau}\rangle-\langle \hat{n}_{l,\tau}\rangle\langle \hat{n}_{m,\tau'}\rangle\Bigr),
\end{equation}
\noindent and
\begin{equation}\label{CSF}
\mathcal{N}(k)=\frac{1}{L}\sum_{l,m=1}^{L}e^{ik(l-m)}\Bigl(\langle \hat{n}_{l} \hat{n}_{m}\rangle-\langle \hat{n}_{l}\rangle\langle \hat{n}_{m}\rangle\Bigr),
\end{equation}
\noindent where $\langle\cdots\rangle\equiv\langle\psi_0|\cdots|\psi_0\rangle$, with $|\psi_0\rangle$ being the ground-state wave function.\par 
The Luttinger liquid parameter $K_{\rho}$ is given by:
\begin{equation}\label{LLP}
\mathcal{N}(k \to 0)= \frac{3K_{\rho}}{2\pi}|k|,
\end{equation}
\noindent and this takes at value of one for a non-interacting system, while it is larger (lower) than one for attractive (repulsive) interactions~\cite{Manmana-PRA11}.\par 
\begin{figure}[t] 
\includegraphics[width=21pc]{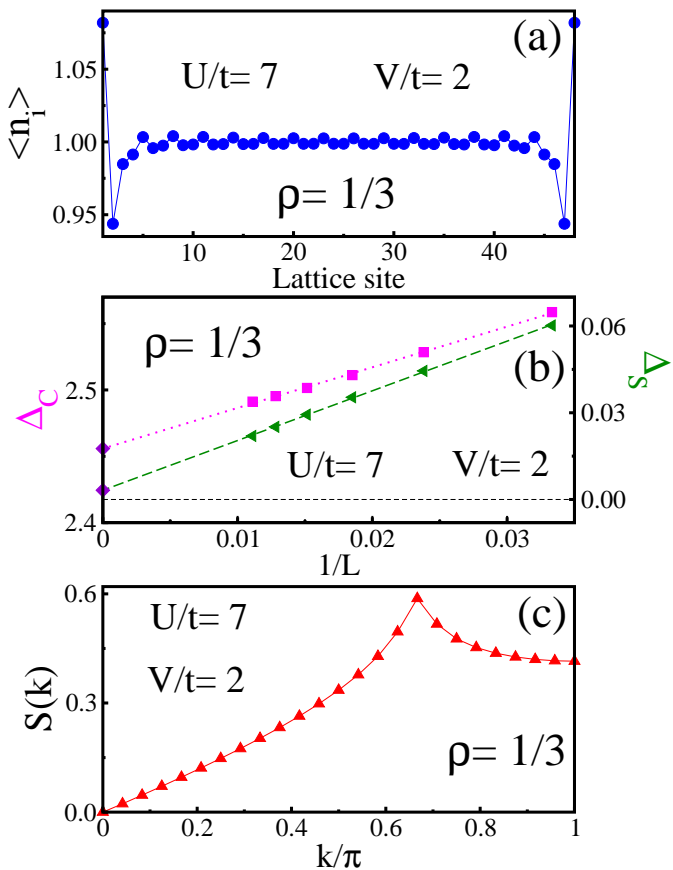} 
\caption{\label{fig2} Physical properties of an extended SU$(3)$ Hubbard model with $U/t=7$ and $V/t=2$. (Top) Local density $ \left \langle n_i \right \rangle$ as a function of local site $i$ for a lattice with $L=48$ sites. The occupation is constant and around one along the lattice, characterizing an SDW phase. The anomalies at the ends correspond to finite boundary effects. (Middle) The charge ($\Delta_C$) and spin ($\Delta_S$) gap as a function of the inverse of the lattice size $1/L$. The extrapolated value at the thermodynamic limit is represented by a diamond. (Bottom) Spin structure factor $\mathcal{S}(k)$ for a finite lattice of size $L=48$. Clearly, a maximum at $k=2\pi/3$ can be seen, this being a fingerprint of the SDW state~\cite{Manmana-PRA11}. In all plots, the points correspond to DMRG results and the lines are visual guides.}
\end{figure} 
\section{\label{sec3}  Results }
We will explore the ground state of the system by sweeping through positive and negative values of the local and next-neighbor interactions. Calculating the properties defined above, we characterize the various phases of the system, which are shown in Fig.~\ref{fig1}. We highlight that the frontiers between the different phases were determined using the tools of quantum information theory, namely the von Neumann block entropy and the fidelity. We work with a total number of particles equal to the lattice size, i.e. at a global density of $\rho=1/3$.\par 
\subsection{Spin-density wave (SDW) state}
Repulsive local and next-neighbor interactions prevent the accumulation of carriers at a single site, and considering that the total number of particles  matches the lattice size, it is expected that in some regime of parameters the lower energy configuration will be that where one carrier occupies a single site. This happens exactly as shown in Fig.~\ref{fig2} (a) for $U/t=7$ and $V/t=2$, where the expectation value of the local density operator $ \left \langle \hat{n}_i \right \rangle \approx1$, except on the borders, due to the open boundaries conditions.\par 
In Fig.~\ref{fig2} (b), we show the evolution of the charge and spin gap as the lattice size increases, observing that they decrease monotonously. Fitting the curve to the expression $\Delta_C(L) = \Delta_C + a/L + b/L^2$, with $\Delta_C$, $a$, and $b$ free parameters, we obtained a finite charge gap at the thermodynamic limit $\Delta_C/t=2.46$. Repeating the above procedure for the spin gap, we observed that the spin gap is very small at the thermodynamic limit and we treat it as zero.\par
Note that for $V=0$, a quantum phase transition from a metallic to a Mott  insulator state was found for a finite value of the local interaction~\cite{Buchta-PRB07,Manmana-PRA11}. This insulator state with zero spin gap and finite charge gap is what we have obtained here for a nonzero value of $V$. Another fingerprint of this spin-density wave state is shown in  Fig.~\ref{fig2} (c), where the spin structure factor appears, which exhibits a maximum at the wavevector $k=\tfrac{2}{3}\pi$, a fact that characterizes this state, as shown by Manmana {\it et al.}~\cite{Manmana-PRA11}.\par 
\begin{figure}[t] 
\includegraphics[width=19pc]{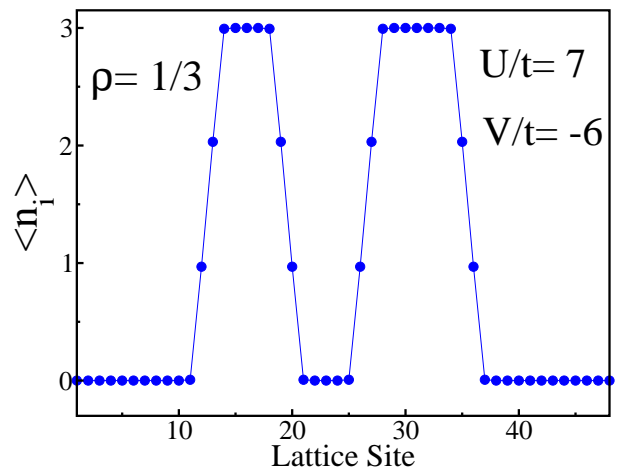} 
\caption{\label{fig3} Occupation ($ \left \langle n_i \right \rangle$) pattern along the lattice for an extended SU$(3)$ Hubbard model with 
$U/t=7$ and $V/t=-6$. Here, the lattice size  is $L=48$, and domains with sites fully occupied and others with empty sites can clearly be seen, suggesting a phase separation (PS) state. The symbols are DMRG results and the lines are visual guides.}
\end{figure} 
\subsection{Phase separation (PS) state}
Considering local attractive interactions ($U<0$), we expect the carriers to try to group themselves at some sites, leaving others empty, due to the density considered here. Adding an attractive nonlocal interaction ($V<0$) will lead to the formation of domains of occupied and empty sites, which can be seen in the density profile shown in Fig.~\ref{fig3} for $U/t=7$ and $V/t=-6$. Note that the form of this profile with domains of fully occupied sites ($ \left \langle \hat{n}_i \right \rangle\simeq3$) and others with empty sites can change with the parameters, this being evidence of the high degeneracy of this state. It is clear that the local compressibility $\kappa_i=\beta \bigl( \langle \hat{n}^{2}_{i}\rangle-\langle \hat{n}_{i}\rangle^{2}\bigr)$ vanishes throughout the lattice, due to the absence of charge fluctuations. However, the description of a phase separation state requires the calculation of the inverse of the global  compressibility $\kappa^{-1}(\rho)=\rho^2\bigl(\partial^2 e(\rho)/\partial \rho^2\bigr)$, where $e(\rho)=E/L$ is the energy density per site~\cite{Moreno-PRB11,Manmana-PRA17}. The inverse of the global compressibility must vanish in a phase separation state. Our results for finite lattices indicate that this will be the case when a calculation at the thermodynamic limit will be done considering large lattice sizes. We do not consider this criterion to determine the borders of this phase, because it numerically is very expensive. \par
\begin{figure}[t] 
\includegraphics[width=20pc]{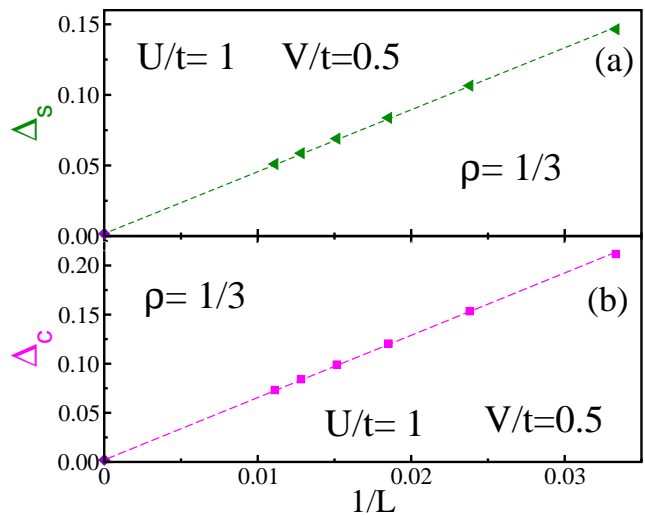} 
\caption{\label{fig4} System size dependence on the spin and charge gaps for a system with interaction parameters $U/t=1$ and $V/t=0.5$. We observe that both quantities tend to zero at the thermodynamic limit. This fact shows that for non-zero values of the next-neighbor interaction it will be a metallic phase. The symbols are DMRG results, and the lines are visual guides.}
\end{figure} 
In Fig.~\ref{fig3}, we see that the phase separation state survives for repulsive values of the local interaction, showing that this phase will dominate the phase diagram for negative values of the next-neighbor interactions in the same way as what happens in the extended $SU(2)$ Hubbard model.\par
This phase separation state has larger charge and spin gaps, which coincide at  the thermodynamic limit, for instance $\Delta _C/t=\Delta _S/t=10.00$ for $U/t=V/t=5$.\par 
We found that for repulsive local interactions and $V<0$, the ground state can be a phase separation one, while for $V\geq0$, it can be a spin-density wave state; hence a quantum phase transition from phase separation to spin-density wave states will take place for negative values of the next-neighbor interactions, as in the $SU(2)$ case.\par 
\subsection{Metallic state}
It is well known that the SU($N$) Hubbard model with $N>2$ undergoes a quantum phase transition from a metallic (gapless) to an insulator (gapped) phase for a finite value of the local interaction. In this paper, we enrich the model by considering a next-neighbor interaction between three-color fermions, and the question that arises is whether the metallic phase survives for a finite value of $V$. In Fig.~\ref{fig4}, we display the charge and spin gaps as a function of the inverse of the lattice size for a system with $U/t=1$ and $V/t=0.5$, and the charge and spin gap decreases linearly as $L$ grows, leading to a very small values at the thermodynamic limit, which will be interpreted as zero. Therefore, the metallic phase will occupy a domain in the phase diagram of the extended SU$(3)$ Hubbard model for repulsive values of $V$.\par 
\begin{figure}[t] 
\includegraphics[width=20pc]{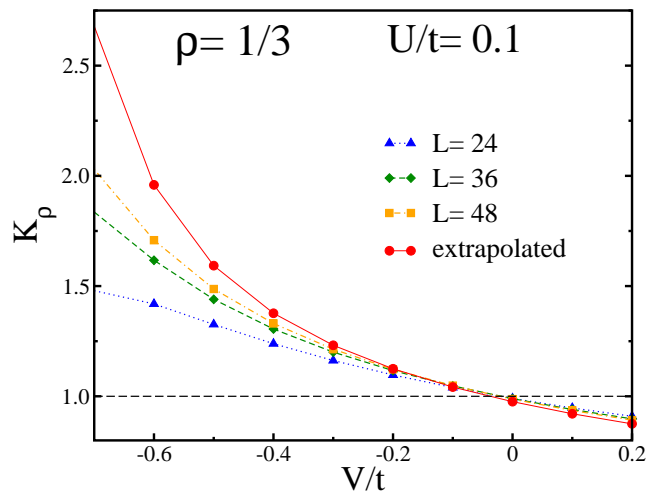} 
\caption{\label{fig5} Luttinger liquid parameter as a function of the next-neighbor interaction for a local interaction $U/t=0.1$. We show the evolution of $K_{\rho}$ for some finite lattices as well as the values extrapolated to the thermodynamic limit (red circles). The lines are visual guides.}
\end{figure} 
\subsection{Pairing phase}
In the absence of next-neighbor interaction ($V=0$), Manmana {\it et al.} show that the SU$(3)$ Hubbard model exhibits a pairing phase for negative values of the local interaction, i.e., the Luttinger liquid parameter is larger than one for attractive interactions~\cite{Manmana-PRA11}. Also, for  $U<0$, a molecular superfluid phase composed of a bound-state made of three fermions for which $K_{\rho}>\sqrt{3}$ was found by Capponi {\it et al.}~\cite{Capponi-PRA08,JSV-PRB01,JSV-JPCM01,Capponi-AP16}.\par 
In order to establish whether a pairing phase emerges for finite values of the next-neighbor interaction, we calculated the Luttinger liquid parameter as a function of $V$ for a local interaction $U/t=0.1$ (see Fig.~\ref{fig5}).
The Luttinger liquid parameter for finite lattices and its values at the thermodynamic limit decrease monotonously as the next-neighbor interaction grows, a behavior that was observed for other positive and negative values of $U$. Our results show that always $K_{\rho}<3$, fulfilling the condition established by Capponi {\it et al.}~\cite{Capponi-PRA08}. The fact that the Luttinger liquid parameter is larger than one suggests a pairing phase, as in the SU$(2)$ model, which will occupy a domain in the phase diagram, due to $K_{\rho}>1$ from $V/t=-0.7$ to $V/t<-0.04$ in Fig.~\ref{fig5}. The latter  value determines the border for which the non-interacting case ($K_{\rho}=1$) is reached, after which a new state emerges, this being a criterion for finding the upper border of the pairing phase.\par 
For lower values of $V$, the Luttinger liquid parameter vanishes, indicating the presence of another phase, in this case the phase separation one; however, due to the irregular behavior of $K_{\rho}$ around the transition, we could not show the overall evolution in Fig.~\ref{fig5}.\par 
\begin{figure}[t] 
\includegraphics[width=20pc]{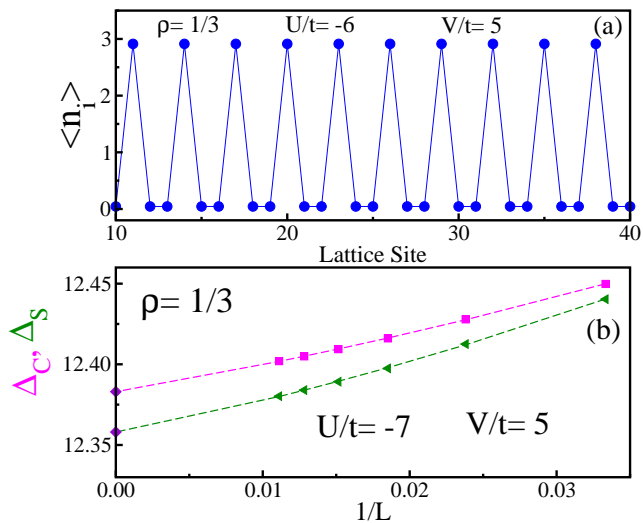} 
\caption{\label{fig6} (a) Density profile along the lattice for a system with interaction parameters $U/t=-6$ and $V/t=5$. The lattice size is $L=48$, and a unit cell with three sites, one fully occupied and the others empty, is clearly seen. This is a charge-density wave (CDW-I) state. (b) Charge (magenta squares) and spin (green  triangle left) gaps as a function of the inverse of the lattice size $1/L$ for $U/t=-7$ and $V/t=5$. At the thermodynamic limit, finite values of charge and spin gaps (diamond points) were obtained. The symbols correspond to DMRG results, and the lines are visual guides.}
\end{figure} 
\subsection{Charge-density wave states}
The attractive local interaction and repulsive next-neighbor interaction interplay should lead to a configuration of fully  isolated  sites; for instance, in the $SU(2)$ case, we expected the following distribution of particles: $\{\ldots,\uparrow\downarrow,\text{ },\uparrow\downarrow,\text{ },\uparrow\downarrow,\text{ },\ldots\}$, i.e. a unit cell with two sites composed of doublons and an empty site.\par 
\begin{figure}[t] 
\includegraphics[width=20pc]{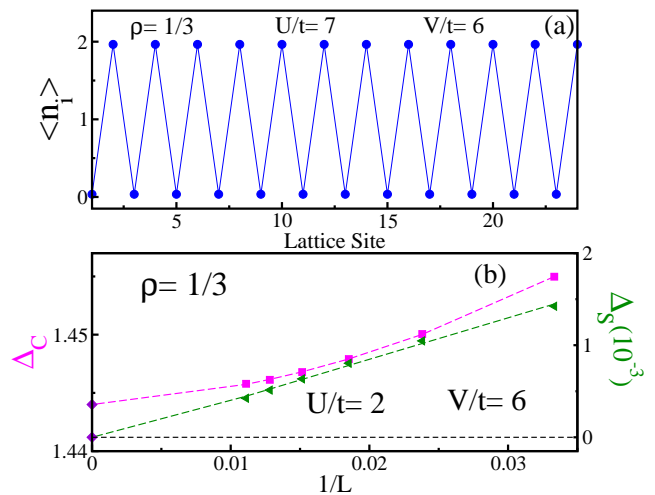} 
\caption{\label{fig7} (Top) Local density $ \left \langle n_i \right \rangle$ as a function of local site $i$ for a lattice with $L=24$ sites and periodic boundary conditions. Here $U/t=7$ and $V/t=6$. We observe a different modulation of charge along the lattice, which will be our CDW-II phase. (Bottom) System size dependence on the charge and spin gaps for a system with interaction parameters $U/t=2$ and $V/t=6$. The finite value for $1/L= 0$ (diamond) corresponds to an extrapolation to the thermodynamic limit, for which the spin gap is very small. In all the  plots, the points correspond to DMRG results and the lines are visual guides.}
\end{figure} 
For a three-color fermion system, the local density profile for $U/t=-6.0$ and $V/t=5.0$ is shown in Fig.~\ref{fig6} (a). The ground state exhibits a periodic modulation of charge, characterized by a unit cell with three sites, two of which are empty and the other full. Therefore, many singlet molecules of three atoms organize themselves into a period-3 crystalline structure. From  this point on, we refer to this state as a charge density wave (CDW-I).\par 
The evolution of the charge and spin gaps versus the inverse of the lattice
length for $U/t=-7.0$ and $V/t=5.0$ is shown  in Fig.~\ref{fig6} (b). Again we observe that these quantities decrease monotonously as the lattice size increases, and fitting the curves with a second-order polynomial function, finite values $\Delta _C/t=12.38$ and $\Delta _S/t=12.36$ at the thermodynamic limit for both gaps were obtained.\par 
According to the previous discussion, for some attractive local interactions the ground state passes from a phase separation state for negative values of $V$ to a (CDW-I) state for positive values of $V$, which indicates that the next-neighbor interaction will drive quantum phase transitions in the system.\par 
For large positive values of local and next-neighbor interactions, we expected  modulations of the local number of carriers throughout the lattice, which effectively happens, as can be seen in Fig.~\ref{fig7}(a). The latter figure was obtained using periodic boundary conditions in order to avoid border effects; a lattice size of $L=24$ sites and $4000$ DMRG's states were considered. Note that the charge modulation obtained differs from that discussed above. Now we do not have a three-site unit cell; instead, the density profile exhibits a two-site unit cell, one empty and the other with $\left \langle n_i \right \rangle \approx 2$, where all colors contribute equally. This state will be our CDW-II state, for which we obtained that the spin gap at the thermodynamic limit is very small and can be interpreted as zero, while the charge gap is finite, for instance $\Delta_C/t=1.44 $ for $U/t=2.0$ and $V/t=6.0$ (see Fig.~\ref{fig7}(b)).\par  
The above discussions allow us to conclude that the SU$(3)$ extended Hubbard model exhibits two charge-density wave phases, in contrast to the single charge-density wave phase reported for the SU$(2)$ case. The (CDW-I) phase dominates for negative values of the local interaction, while the (CDW-II) phase dominates for positive values of $U$.\par
Figures ~\ref{fig2}(a) and ~\ref{fig7}(a) show that fixing the local repulsion ($U/t=7$), the ground state can evolve from a SDW for $V/t=2$ to a (CDW-II) state for $V/t=6$; therefore, there will be a quantum phase transition between these phases.\par 
\begin{figure}[t] 
\includegraphics[width=20pc]{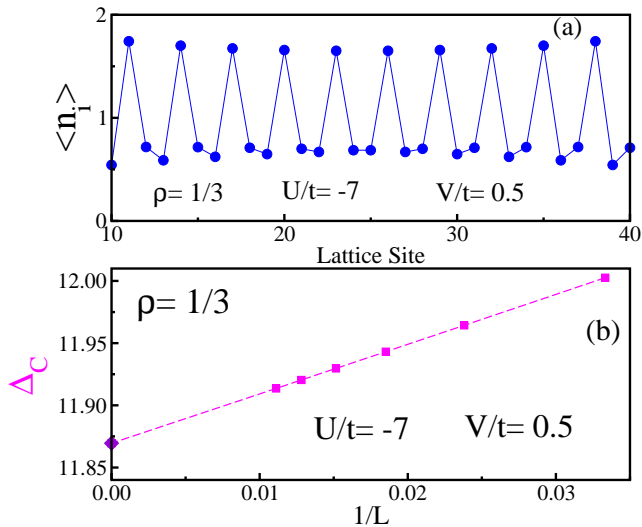} 
\caption{\label{fig8} (Top) Occupation ($ \left \langle n_i \right \rangle$) pattern along the lattice for a lattice with $L=48$ sites and interaction parameters $U/t= -7$ and $V/t= 0.5$. Clearly, there is a modulation of charge with a three-site unit cell, but its extreme values change with the next-neighbor interaction. (Bottom) Charge gap as a function of the inverse of the lattice size $1/L$ for $U/t=-7$ and $V/t=0.5$. The diamond point represents the value at the thermodynamic limit. The spin gap vanishes in this phase. The symbols correspond to DMRG results, and the lines are visual guides.}
\end{figure} 
\subsection{Intermediate phase}
For large attractive local interactions, we observe that  the ground-state can be a phase separate or CDW-I for attractive or repulsive next-neighbor interactions, respectively. Here the question arises if the transition between the above phases is direct, as in the SU$(2)$ case, or if there is another phase between them. We explored this and found a phase with a modulation of charge with a three-site unit cell (see Fig.~\ref{fig8} (a)), whose amplitude increases progressively with the next-neighbor interaction, but the expectation values of the local density do not take the values zero or three, establishing an essential difference between this intermediate phase and the CDW-I one. In Fig.~\ref{fig8} (b) a charge gap appears for a chain with parameters $U/t=-7$ and $V/t=0.5$, showing that at the thermodynamic limit the charge gap is finite but lower than the one found in the CDW-I phase with $U/t=-7$ and $V/t=5$ (Fig.~\ref{fig6} (b)). For a different set of parameters, 
we established that at the thermodynamic limit the spin gap is zero (not shown), which agrees with the fact that the model could  be mapped onto an effective spinless fermionic chain with very small hopping. In summary, the intermediate phase exhibits a modulation of charge without spin gap and a finite charge gap.\par
\subsection{Phase diagram}
\begin{figure}[t!]
\begin{minipage}{18.9pc}
\includegraphics[width=18.9pc]{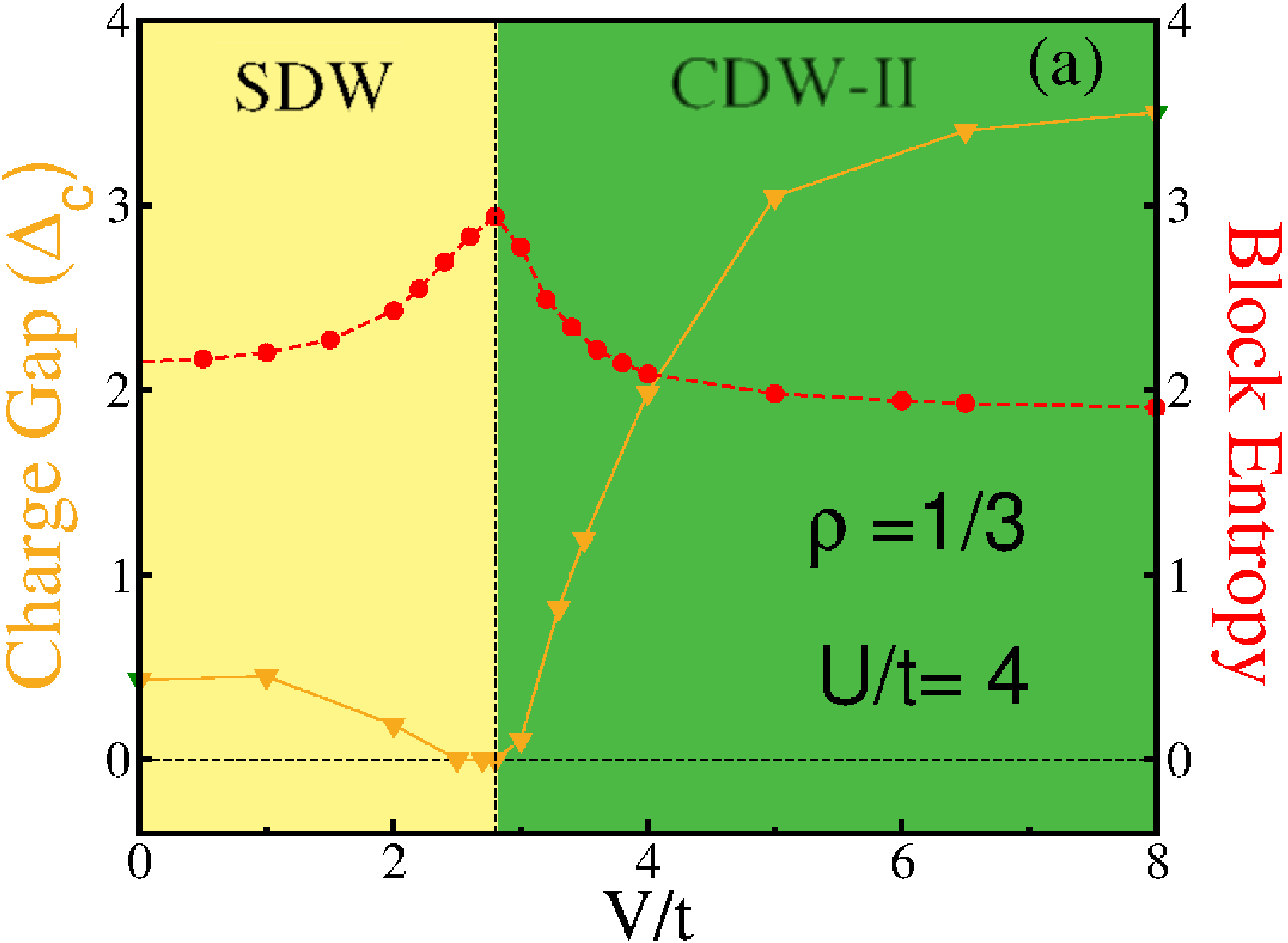}
\end{minipage}
\hspace{5pc}%
\begin{minipage}{18.9pc}
\includegraphics[width=18.9pc]{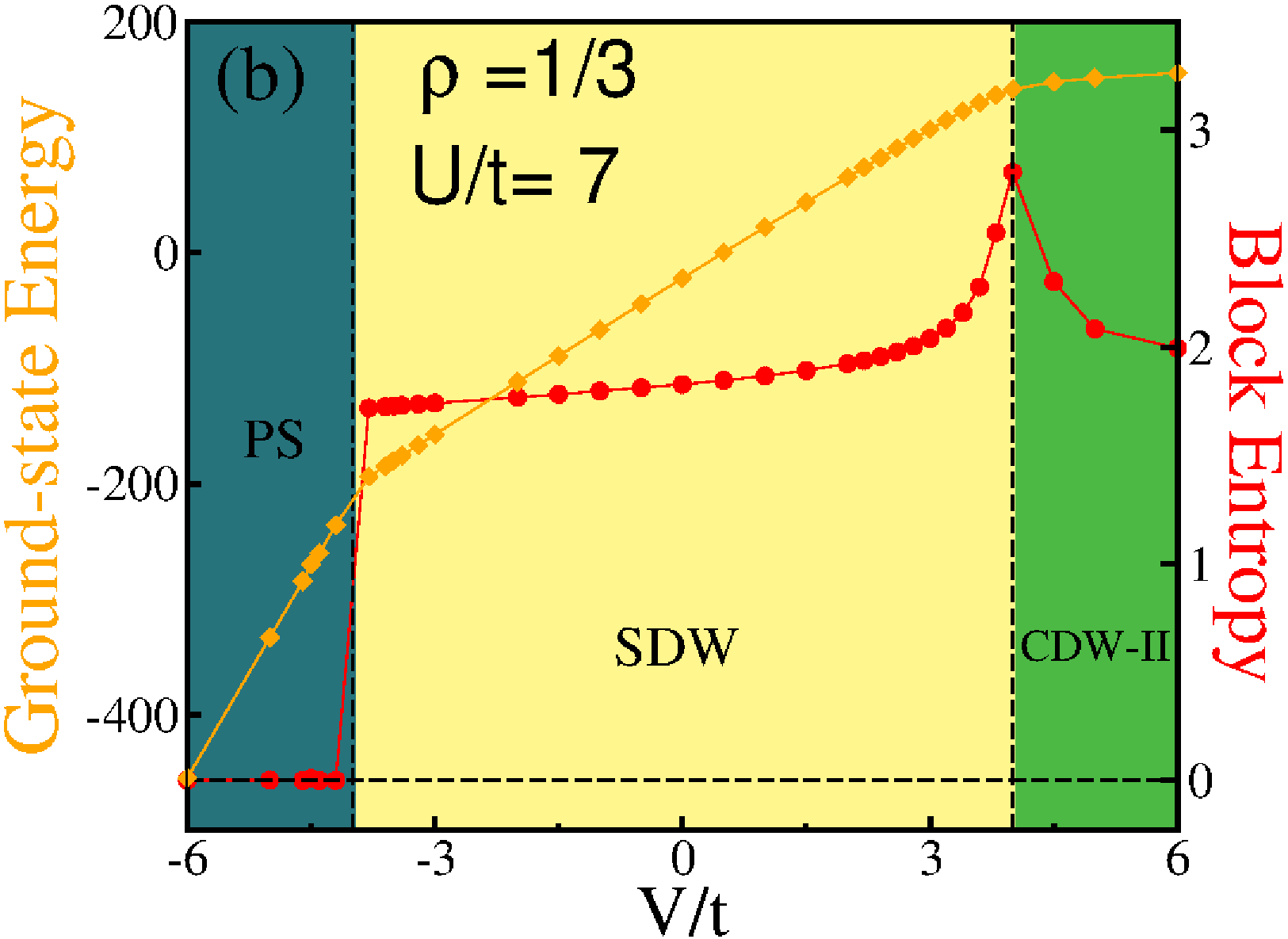}
\end{minipage}
\hspace{5pc}%
\begin{minipage}{18.9pc}
\includegraphics[width=18.9pc]{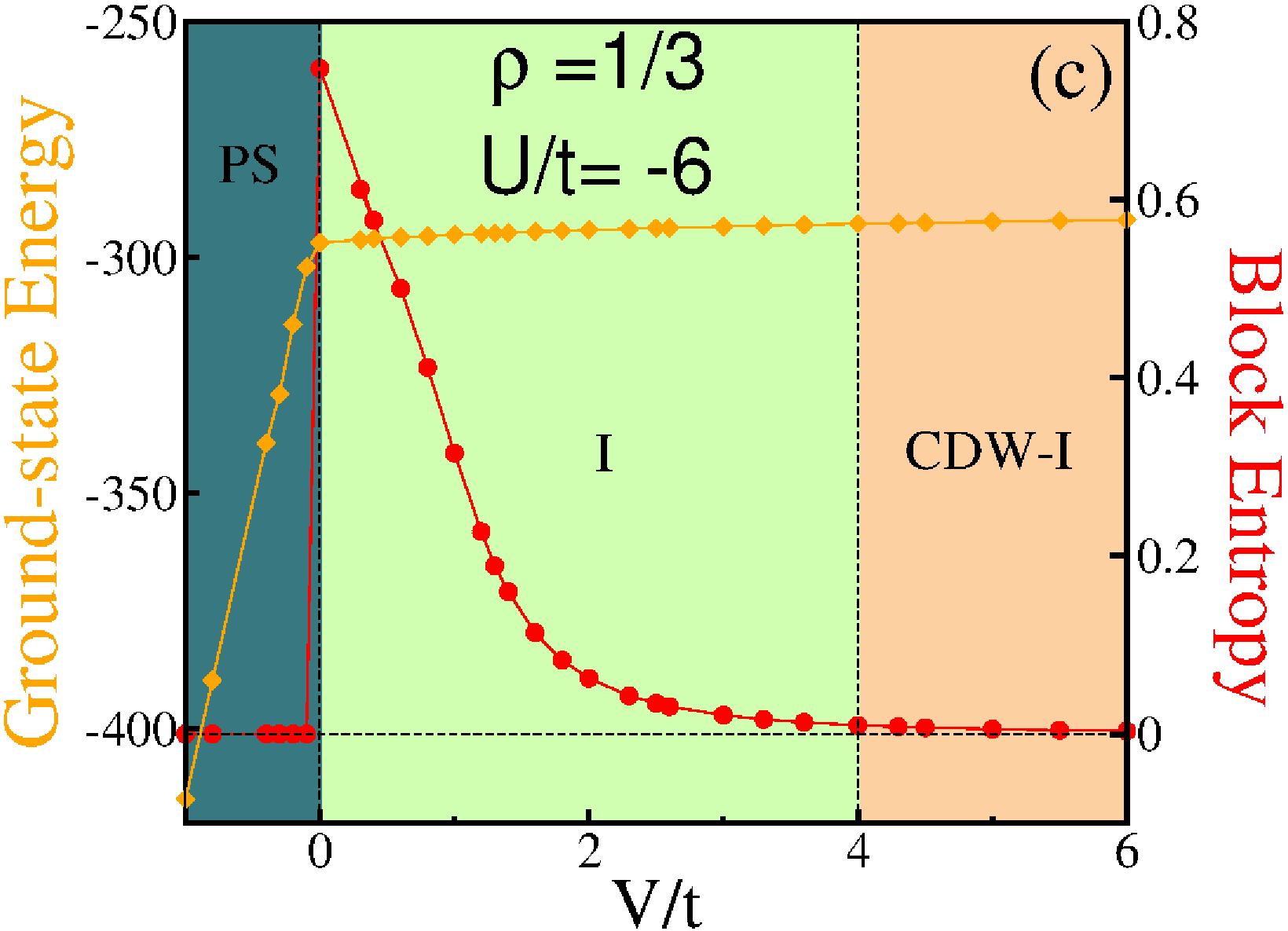}
\caption{\label{fig9} Charge gap, ground-state energy and von Neumann block entropy versus the next-neighbor interaction for different fixed values of $U$. In plot (a), we show the charge gap and the block entropy for $U/t=4$, clearly observing  that there are two regions, determined by the critical point $V_c/t= 2.8$. In plot (b), the local repulsion is $U/t=7$, and the ground-state energy and the block entropy suggest two critical points. Also, two phase transitions are suggested in plot (c), where the local coupling is $U/t=-6$. In all plots, the points correspond to DMRG results and the lines are visual guides.}
\end{minipage}\hspace{2pc}%
\end{figure}

We have shown that adding a next-neighbor interaction term to the SU$(3)$ Hubbard model leads us to a new and rich model that exhibits several different phases, which were characterized using the energy gaps and the correlations of the ground state. For instance, we note that the charge gap takes different values in the diverse phases found.  Therefore, this quantity can be used to determine the critical points that separate two phases for a given set of parameters. The evolution of the charge gap at the thermodynamic limit as a function of next-neighbor interaction appears in Fig.~\ref{fig9}(a) for $U/t=4$. In the absence of next-neighbor interaction ($V=0$), the charge gap is finite and the system is in the SDW phase. Increasing $V$, we note that the charge gap initially remains almost constant and then diminishes, until it vanishes at around $V/t\approx2.7 \pm 0.1$. We must remember that this  criterion is not definitive. After that, the charge gap increases very quickly, and the results suggest that it tends to saturate for larger values of the next-neighbor interaction. It is clear that fixing the local interaction at $U/t=4$ and varying the parameter $V$, the system undergoes a quantum phase transition from an SDW phase for $V/t<2.6$ to a (CDW-II) phase for $V/t>2.8$, and $V/t\approx2.7$ is the critical point for the SDW-(CDW-II) transition for $U/t=4$. To determine the other critical points for this transition and the others suggested, we can follow this procedure and also calculate the spin gap.  However, this approach is too expensive to be considered in a model with very slow convergence that needs a huge amount of DMRG states. Therefore, an alternative procedure  for estimating the critical points is necessary.\par 
Entanglement and the different witnesses used to quantify it have become a very useful tool for localizing and studying quantum phase transitions without {\it a priori} knowledge about it~\cite{Amico-RMP08,Laflorencie-PR16}. Also, recently a direct measure of quantum purity, R\'enyi entanglement entropy, and mutual information was done in a one-dimensional Bose-Hubbard system composed of $^{87}Rb$ atoms~\cite{Islam-N15}. A proposal to measure entanglement witnesses in fermionic systems was suggested recently~\cite{Cornfeld-PRA19}.\par
We want to point out that diverse entanglement witnesses have been used to establish the boundaries between the phases in the phase diagram of the extended SU$(2)$ Hubbard model~\cite{Gu-PRL04,Deng-PRB06,Mund-PRB09,Iemini-PRB15,Spalding-PRB19}.\par 
One of the most-used entanglement witnesses is the von Neumann block entropy, which is defined by $ S=-Tr\varrho_A  ln  \varrho_A$, where $\varrho_A=Tr_B \tilde{\varrho}$ is the reduced density matrix of block $A$ with $l$ sites, 
$\tilde{\varrho}= |\Psi\rangle \langle \Psi|$ is the pure-state density matrix of the whole system, $B$ is a block with $L-l$ sites, and  in particular in this paper we consider  $l=L/2$. Quantum phase transitions are signaled by the von Neumann block entropy by means of minimums, maximums, or discontinuities in the entropy and its derivative~\cite{Franca-PRL08,MendozaArenas-PRA10}.\par 
In Fig.~\ref{fig9}(a), we display the von Neumann block entropy for a lattice of $L=48$ sites as a function of the next-neighbor interaction for a local repulsion $U/t=4$. We see that the von Neumann block entropy as a function of $V$ initially grows and then decreases, determining a maximum value of the von Neumann block entropy at $V_c/t\approx2.8$, which divides the figure and makes it unavoidable to think that this maximum corresponds to the critical point despite being slightly displaced from the position found using the charge gap at the thermodynamic limit, which may be due to the difficulty of determining exactly where the charge gap vanishes. Our numerical results indicate that the position of the anomalies in the block entropy does not depend on the lattice size, i.e. there are no finite size effects in the localization of the critical points. Therefore, we conclude that the von Neumann block entropy is useful for estimating the critical points of the SDW-(CDW-II) transition in the extended SU$(3)$ Hubbard model, and we determine all the diamond points in Fig.~\ref{fig1} in this way.\par
The evolution of the von Neumann block entropy versus the next-neighbor interaction for $U/t=7$ is shown in Fig.~\ref{fig9}(b). For large negative values of $V$, the system will be in the phase separation state, for which the ground state can be expressed as a product of the states of a single site, i.e. this a separable state; therefore, the entanglement must be canceled, which we effectively observe in the von Neumann block entropy. But in $V/t\approx-3.8$, the von Neumann block entropy jumps, indicating that the PS-SDW transition has occurred. Inside the SDW phase, the block entropy remains almost constant, reflecting the small variation in the number of degrees of freedom that is associated with the fact that there is one  particle per site. Around a quantum phase transition, fluctuations in the number of degrees of freedom are expected, which can be indicated by the entanglement. Effectively, we see that the von Neumann block entropy increases quickly, reaches a maximum, and then decreases quickly. This happens around $V/t\approx 3.7$ and determines the quantum phase transition from the SDW phase to the (CDW-II) one, in the same manner as discussed in the previous figure. Comparing the critical points obtained for the SDW-(CDW-II) transition, we conclude that the values of the critical couplings will increase with the local repulsion. Our numerical results suggest that for large positive values of $V$, the von Neumann block entropy tends to be constant, which reflects the fact that the ground state will not change. We conclude that the von Neumann block entropy allows us to estimate the critical points of the PS-SDW and SDW-(CDW-II) quantum phase transitions. To characterize the quantum phase transitions discussed above, we show in Fig.~\ref{fig9}(b) the ground-state energy versus $V$ (orange diamonds), and clearly observe a change in the slope of the curve around the critical points, which indicates that both transitions are of the first order. Replicating the above procedure for other positive values of $U$, we estimate the critical points for the SDW-(CDW-II) transition (diamond points) and the PS-SDW transition (black circles) shown in the phase diagram (Fig.~\ref{fig1}).\par 
Fixing the local repulsion at $U/t=-6$ and varying the next-neighbor interaction, we calculate the von Neumann block entropy, obtaining the results shown in Fig.~\ref{fig9}(c). For the phase separation region, the block entropy is again zero, but now the discontinuity takes place at $V/t=0$, signaling the PS-intermediate quantum phase transition, due to  the huge increase in the number of degrees of freedom when the system passes from the PS phase to the intermediate one. After this discontinuity and increasing the next-neighbor interactions, the block entropy diminishes  monotonously, reflecting the tendency to generate a definitive modulation, and it vanishes from $V/t\approx 4$, marking the beginning of the CDW-I phase, whose ground state is separable. Hence in the second quadrant of the phase diagram, we use the cancellation of the block entropy as a criterion for determining the critical point for the intermediate-(CDW-I) transition (triangle up points in Fig.~\ref{fig1}). Again, we displayed the evolution of the ground-state energy as a function of $V$ and found an evident change of slope around the PS-intermediate transition, which indicates that this transition is of the first order. For larger values of $V$, the ground-state energy increases monotonously and very slowly, indicating that the intermediate-(CDW-I) transition is continuous.\par
\begin{figure}[t] 
\includegraphics[width=20pc]{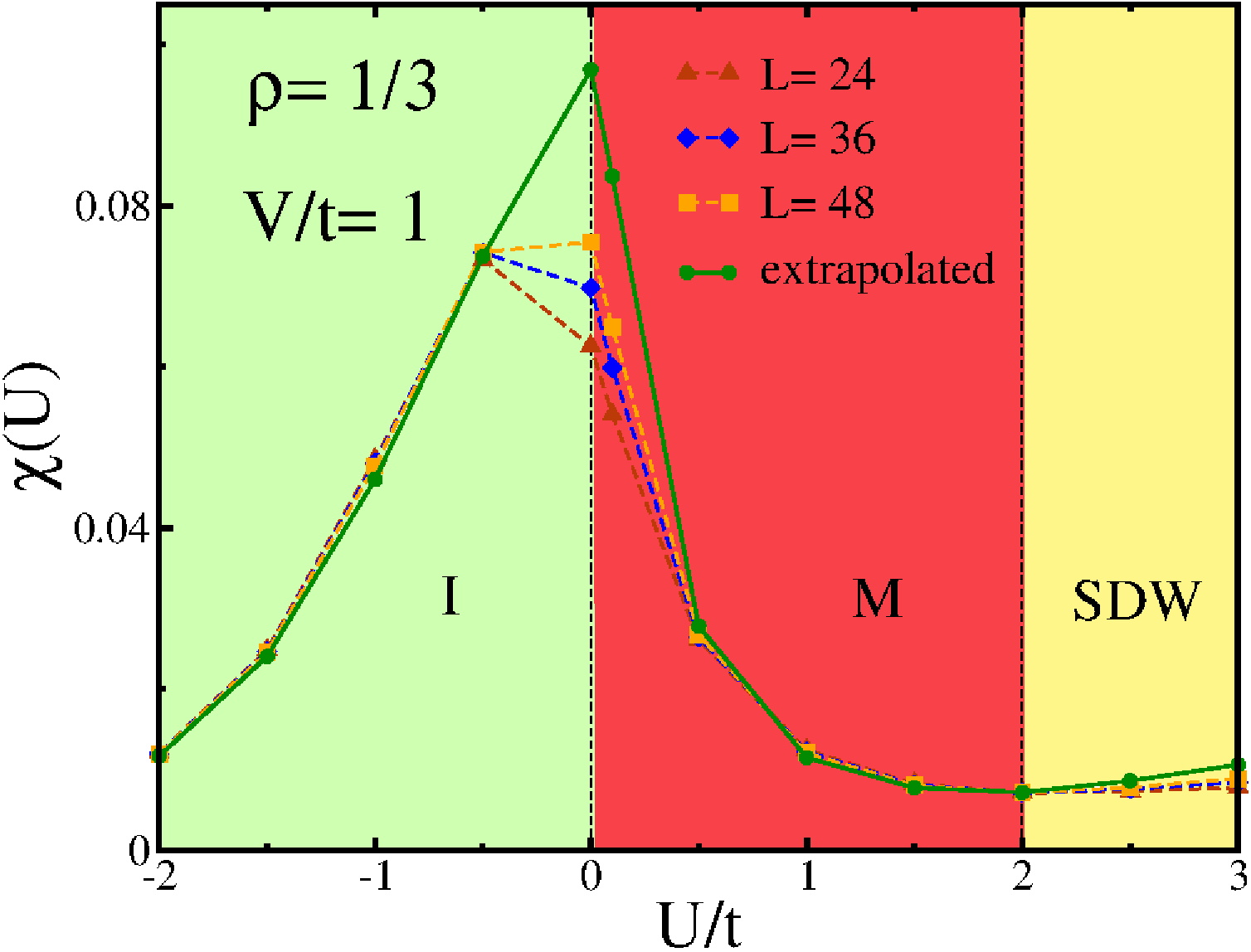} 
\caption{\label{fig10} Fidelity susceptibility as a function of the local coupling with $V/t =1$. At the thermodynamic limit, the fidelity susceptibility exhibits extreme values, which suggests quantum phase transitions. The symbols correspond to DMRG results, and the lines are visual guides.}
\end{figure} 
To determine the borders that define the metallic domain in the phase diagram, we used the fidelity 
\begin{equation}
 \mathcal{F}(U)=\lvert \langle \psi_0 (U)| \psi_0 (U+dU)\rangle \rvert,
\end{equation}
\noindent and the fidelity susceptibility
\begin{equation}
  \chi(U)=\frac{2\bigl[1-\mathcal{F}(U) \bigr]}{LdU^2},
\end{equation}
\noindent which were used successfully in the case of $V= 0$~\cite{Manmana-PRA11}.\par 
In Fig.~\ref{fig10}, the fidelity susceptibility appears as a function of the local interaction for $V/t= 1$. We display the evolution of the fidelity susceptibility for some finite lattice sizes, showing diverse behavior, depending on $U$. The green circles correspond to the values at the thermodynamic limit. Starting from negative values of $U$, i.e. inside of the intermediate phase, the  fidelity susceptibility grows monotonously, reaching a maximum at $U/t= 0$. This fact and the ground state exploration around this local interaction value allow us to conclude that the intermediate-metallic transition takes place at this value, a scenario that is repeated for larger values of $V$. For positive values of $U$, the fidelity susceptibility decreases in a way similar to that without next-neighbor interaction~\cite{Manmana-PRA11} and reaches a minimum around $U/t=2$, indicating the passage from the metallic phase to the SDW one. Considering other values of $V$ and using the fidelity susceptibility, we determine the critical points (squares in Fig.~\ref{fig1}) that separate the metallic phase from the SDW or CDW-II phases. Using the evolution of the ground-state energy around the critical points, we conclude that the transitions to and from the metallic phase are continuous.\par 
It is important to clarify that except for the top border of the pairing phase, all the borders of the phase diagram (Fig.~\ref{fig1}) were obtained using the anomalies in the von Neumann block entropy or the fidelity susceptibility.\par 
Comparing the SU$(3)$ phase diagram with its SU$(2)$ counterpart, the main differences are the emergence of an intermediate phase, a metallic one, and two charge-density wave phases in the phase diagram for three-color fermions.\par
\section{\label{sec4} Conclusions}
We have studied an extended SU$(3)$ Hubbard model, i.e. a three-color fermion gas where the carriers interact locally and with next-neighbors. Using the density matrix renormalization group technique and the DBSS protocol, we calculated the spin and charge gaps, the spin and charge structure factors, the Luttinger liquid parameter, the density profiles, the von Neumann block entropy, and the fidelity susceptibility.\par 
In the same way as its SU$(2)$ counterpart, the model considered exhibits a rich phase diagram at a global filling of one particle per site. As expected,  an SDW phase emerges with a homogeneous distribution of charge (each site occupied by one fermion), zero spin gap, and a maximum in the spin structure factor at $k=2\pi/3$. Also, for negative next-neighbor interaction there arises a phase separation state, characterized by domains with full (three particles) or empty sites. Another separable state that emerges is the charge density wave (CDW-I), which exhibits a modulation of charge along the lattice with a unit cell of three sites, where one site is full and the others empty. A domain with pairing between the carriers was identified, although its specific characteristics will be studied later. The above  phases are in some way similar to those obtained in the SU$(2)$model, but for repulsive next-neighbor interaction, new phases arise, such as a metallic phase, a charge density wave (CDW-II) with a two site unit cell, and finally an intermediate phase, which is a non-separable state with modulation of charge.\par
We observed that the von Neumann block entropy is zero in the PS and CDW-I states,  because these states are separable. On the other hand, a constant non-zero value was found for the SDW and the CDW-II states. A discontinuity from zero to finite value in the block entropy signals the quantum phase transitions that involve the phase separation state. Meanwhile, a maximum in the von Neumann block entropy determines the critical point that separates the SDW and the CDW-II phases. Also, we note that a maximum and a minimum in the fidelity susceptibility mark the intermediate-metallic and metallic-SDW transitions, respectively. We largely relied on these characteristics of the evolution of the block entropy and the fidelity susceptibility in order to build the phase diagram of the extended SU$(3)$ Hubbard model, which was completed by  calculating the Luttinger liquid parameter, obtaining Fig.~\ref{fig1}.\par 
Taking into account that recently $^6$Li atoms undergoing non-local interactions in a two-dimensional optical lattice were studied in experiments~\cite{GuardadoS-Arxiv20}, we believe that the results reported here can stimulate additional experimental efforts involving Rydberg atoms and more internal degrees of freedom.\par 
\section*{Acknowledgments}
J.S.-V. acknowledges helpful discussions with J. J. Mendoza-Arenas. The authors are thankful to {\"O}rs Legeza, who provided data for comparison. This research was supported by DIEB- Universidad Nacional de Colombia (Grant No. 51116).

%
\bibliography{/home/jereson/PAPERS/Bib/Bibliografia.bib}

\end{document}